\documentclass[prd,superscriptaddress,preprint]{revtex4}

\usepackage{mathrsfs,amsmath}
\usepackage{eurosym}
\usepackage{amssymb}
\usepackage{bbold}

\usepackage{latexsym,epsfig,epstopdf}
\usepackage{amssymb,amsmath,amsthm}

\usepackage{times}
\usepackage{color}

\usepackage{lmodern}
\usepackage[utf8]{inputenc}

\usepackage{graphicx}

\usepackage{color}

\usepackage{braket}

\usepackage[mathscr]{euscript} 

\def\be{\begin{equation}}
\def\ee{\end{equation}}
\def\bea{\begin{eqnarray}}
\def\eea{\end{eqnarray}}

\begin{document}

\title{Topological invariants of complex energy plane in non-Hermitian systems}
\author{Annan Fan}
\affiliation{School of Physics, Sun Yat-Sen University, Guangzhou, 510275, China}
\author{Shi-Dong Liang }
\altaffiliation{Email: stslsd@mail.sysu.edu.cn}
\affiliation{School of Physics, Sun Yat-Sen University, Guangzhou, 510275, China}
\affiliation{State Key Laboratory of Optoelectronic Material and Technology, and\\
Guangdong Province Key Laboratory of Display Material and Technology,
Sun Yat-Sen University, Guangzhou, 510275, China}

\date{\today }
\begin{abstract}
Non-Hermitian systems as theoretical models of open or dissipative systems exhibit rich novel physical properties and fundamental issues in condensed matter physics.We propose a generalized local-global correspondence between the pseudo-boundary states in the complex energy plane and topological invariants of quantum states. We find that the patterns of the pseudo-boundary states in the complex energy plane mapped to the Brillouin zone are topological invariants against the parameter deformation. We demonstrate this approach by the non-Hermitian Chern insulator model. We give the consistent topological phases obtained from the Chern number and vorticity. We also find some novel topological invariants embedded in the topological phases of the Chern insulator model, which enrich the phase diagram of the non-Hermitian Chern insulators model beyond that predicted by the Chern number and vorticity. We also propose a generalized vorticity and its flipping index to understand physics behind this novel local-global correspondence and discuss the relationships between the local-global correspondence and the Chern number as well as the transformation between the Brillouin zone and the complex energy plane. These novel approaches provide insights to how topological invariants may be obtained from local information as well as the global property of quantum states, which is expected to be applicable in more generic non-Hermitian systems.
\end{abstract}

\maketitle



\section{Introduction}

Non-Hermitian quantum systems have attracted growing attention due to their novel physical properties beyond Hermitian systems and potential applications in quantum computation.\cite{Ramy,Gong,Kohei,He} In general, non-Hermiticity appears in open or dissipative systems, such as energy gain or loss, or non-conservation of probability associated with  currents.\cite{Ramy,Liang,Chernyak,Qi} The non-Hermitian systems contain complex eigen energies and nonorthogonal eigenstates, which motivate many attempts to explore their novel physical phenomena and potential applications beyond conventional Hermitian systems, and to explore what novel mathematical structures exist beyond those of canonical quantum mechanics.\cite{Kohei,Ali} A lot of efforts have been focused on the classification of the topological equivalence classes based on the symmetry of systems in condensed matter physics and on the physical properties of these topological phases, as well as their potential applications.\cite{Gong,Kohei,Yu,Annan,Niu,Yang}

The rich structures of complex energy band gaps and complex exceptional points, such as the robust point-gap and line-gap structures, lead to some novel phenomena beyond Hermitian systems.\cite{Kohei} It has been found that non-Hermiticity enriches topological phases beyond those of Hermitian systems. The 10-fold topological equivalence classes of Hermitian systems based on the Altland-Zirnbauer(AZ) symmetry classification were extended to 38-fold topological equivalence classes due to additional sublattice symmetry and pseudo-Hermticity. \cite{Gong,Kohei} These topological equivalent phases of quantum systems can be characterized by topological quantum numbers, such as the winding number, Chern number and vorticity. \cite{Kohei2,Ghatak} The vorticity defined by the complex angle of the complex energy band and has been shown to be equivalent to the winding number classifying the topological invariants for Hermitian systems. \cite{Kohei,Fu}
The robust topological phases are protected by the symmetries of the system. The energy band gap structure and symmetries can be remained under a unitary flattening and Hermitianization of these systems. \cite{Kohei} Even though the non-Hermiticity of systems deforms the Bloch-wave behavior leading to the skin effect for lattice models,\cite{Lee,Alvarez,Kawabata,Liu} the biorthogonal polarization was introduced to modify the conventional bulk-boundary correspondence to detect the zero modes,\cite{Kunst,Yao,Esaki}, the non-trivial edge states and finite-size effects in non-Hermitian systems, such as  the Su-Schrieffer-Heeger (SSH) model.\cite{Zhu,Jiang,Dangel} Thus, the bulk-boundary correspondence is still valid for non-Hermitian systems.\cite{Lieu,Chen,Yin,Rui,Daniel,Flore,Shunyu}
In particular, it has been found that one can construct a non-Hermitian counterpart to any Hermitian system whose long-time dynamics realizes every anomalous Hermitian  boundary mode of the AZ classes, which establishes a correspondence between the topological classifications of (d+1)-dimensional gapped Hermitian systems and d-dimensional point-gapped non-Hermitian systems.\cite{Lee2}

In particular, deviations of the quantized Chern number and the quantum Hall conductance in the non-Hermitian Chern insulator were found due to non-Hermitian effects.\cite{Yu,Zhao} More interestingly, the quantum Hall conductance in the non-Hermitian Dirac model can be generalized to quantum Hall admittance, namely the quantum Hall susceptance emerges, such as quantum Hall capacity and induction in non-Hermitian systems. This could inspire fundamental insights into non-Hermitian systems and lead to potential application.\cite{Annan}

Non-Hermiticity yields unconventional characteristics in physical systems, especially the existence of the complex eigen energies and nonorthogonal eigenstates  highlights some fundamental issues, such as use of the pseudo-Hermitian concept to reformulate the canonical domain of non-Hermitian systems and relativistic non-Hermitian quantum mechanics.\cite{Dorje,Ali}

The topological classification of quantum states is based on the symmetries of quantum systems, their energy gap structures and the bulk-boundary correspondence.\cite{Gong,Kohei,Zhang}
In particular, the bulk-boundary correspondence uncovers the bulk-boundary duality, in which the edge states play an essential role in the topological phases of systems.
This duality implies that a part of local information in systems dominates some of their global properties. This roots the holographic technology in optics,
and inspires us to explore some way to capture the global topological invariants based on the local information of quantum states in non-Hermitian systems.

The complex energy bands in non-Hermitian systems and their complex band gap structures are mapped to the complex energy plane, which provides a platform to study
the topological invariants in terms of local information in the complex energy plane. The complex exceptional point gap and line gap defects are robust and are associated with the topological phases. \cite{Kohei} This provides some hints about how to search for local states in the complex energy plane in order to detect the topological invariant.

From a mathematical point of view, topology provides a precise tool to study the global behavior of geometric objects. Many different topological indexes, or numbers, are used to characterize different topological invariants for different geometric objects, such as the intersection number, winding number, linking number, Chern number and Euler characteristic. They label the homeomorphic equivalent classes for different geometric objects.\cite{Eber} Physicist realized some quantum states exhibiting novel geometric properties and topological invariants, which can be described by the winding number, Berry phase, Berry curvature and Chern number.\cite{Bohm,Di} Quantum Hall conductance can be expressed in terms of the Chern number, which inspires a lot of attempts to explore the geometric and topological properties of quantum states and their potential applications. The Berry phase and Berry curvature depend on the wave function of the system and are associated with the Chern number for Hermitian systems.\cite{Bohm,Di}
Recently, introduced the wrapping number was as a unified approach to obtain the topological invariants and give the correlation length, universality classes, and scaling laws associated with topological phase transitions in arbitrary dimensions and symmetry classes for Dirac models.\cite{Chen1}

In this paper, we propose a novel approach for characterizing topological invariants in non-Hermitian systems based on the local-global correspondence between the pseudo-boundary states (PBS) in the complex energy plane and topological invariants of the quantum states. We first set up the conceptual and mathematical framework of this correspondence in Sec. II, in which we present the mathematical structure and physical conclusions of the local-global correspondence, how to connect the PBS to the topological invariants. We also give this correspondence and its homotopic continuous map. In Sec. III, we demonstrate the local-global correspondence by repeating the phase diagram of the non-Hermitian Chern insulator model. Firstly, we present the topological invariant patterns in the Brillouin zone (BZ), which are generated from the PBS in the complex energy plane. We then obtain a phase diagram that is consistent with  previous results in order to demonstrate the validity of the local-global correspondence.
Next, we present a novel topological invariant embedded in the phase diagram based on this correspondence. This topological invariant is hidden in the conventional approaches, such as Chern number, winding number and vorticity. \cite{Kohei2,Ghatak}
In Sec. IV, we discuss the physical and mathematical meanings of the topological invariants behind the local-global correspondence and their relationships to the Chern number and vorticity. We study the transformation from the BZ to the complex energy plane and find the Jacobian determinant to be zero for the PBS. This implies that the BZ and the complex energy plane are not one-to-one. Finally, we present our conclusions and outlook in Sec. V. In the Appendix, we give the detailed algorithm of the local-global correspondence and the derivation of the Jacobian determinants of the transformation from the BZ to the complex energy plane.

\section{The local-global correspondence on complex energy plane}

\subsection{The pseudo boundary states in the complex energy plane}
Let us consider a non-Hermitian Hamiltonian $H^{\dag}\neq H$. Suppose that the Hamiltonian is bounded and works in the Brillouin zone (BZ) with a parameter space, $H(\mathbf{k},\lambda)$, where $\mathbf{k}\in BZ^{d}$ is the BZ and $\lambda\in \mathcal{M}^{p}_{\lambda}$ is a set of parameters, in which $d$ is the dimension of the BZ and $p$ is the dimension of the parameter space.
A pair of the eigen equations for a non-Hermitian Hamiltonian and its adjoint operator are given by\cite{Kohei2,Ghatak}
\begin{subequations}\label{HH}
\begin{eqnarray}
H(\mathbf{k},\lambda)|\psi_{n}^{R}(\mathbf{k},\lambda)\rangle=E_{n}(\mathbf{k},\lambda)|\psi_{n}^{R}(\mathbf{k},\lambda)\rangle \\
H^{\dagger}(\mathbf{k},\lambda)|\varphi_{n}^{L}(\mathbf{k},\lambda)\rangle=E^{*}_{n}(k,\lambda)|\varphi_{n}^{L}(\mathbf{k},\lambda)\rangle.
\end{eqnarray}
\end{subequations}
where $R$ and $L$ label the right and left sides of the inner product respectively for convenience to non-Hermitian systems.
Suppose that the Hilbert space for this non-Hermitian system is separable, then the eigen vectors of the Hamiltonian and its Hermitian adjoint consist of
a biorthogonal basis on the Hilbert space,\cite{Ali}
\begin{equation}\label{Orbs}
\langle\varphi_{m}^{L}(\mathbf{k},\lambda)|\psi_{n}^{R}(\mathbf{k},\lambda)\rangle=\delta_{mn}
\end{equation}
where $|\psi_{n}^{R}(\mathbf{k},\lambda)\rangle$ and $\langle\varphi_{n}^{L}(\mathbf{k},\lambda)|$ are the corresponding eigenstates of the Hamiltonian and its Hermitian adjoint. The completeness relation is given by
\begin{equation}\label{CPR1}
\sum_{n}|\psi_{n}^{R}(\mathbf{k},\lambda)\left\rangle\right\langle\varphi_{n}^{L}(\mathbf{k},\lambda)|=I
\end{equation}
where $I$ is the identity matrix.

\begin{figure}[htbp]
	\centering
	\includegraphics[width=0.8\linewidth]{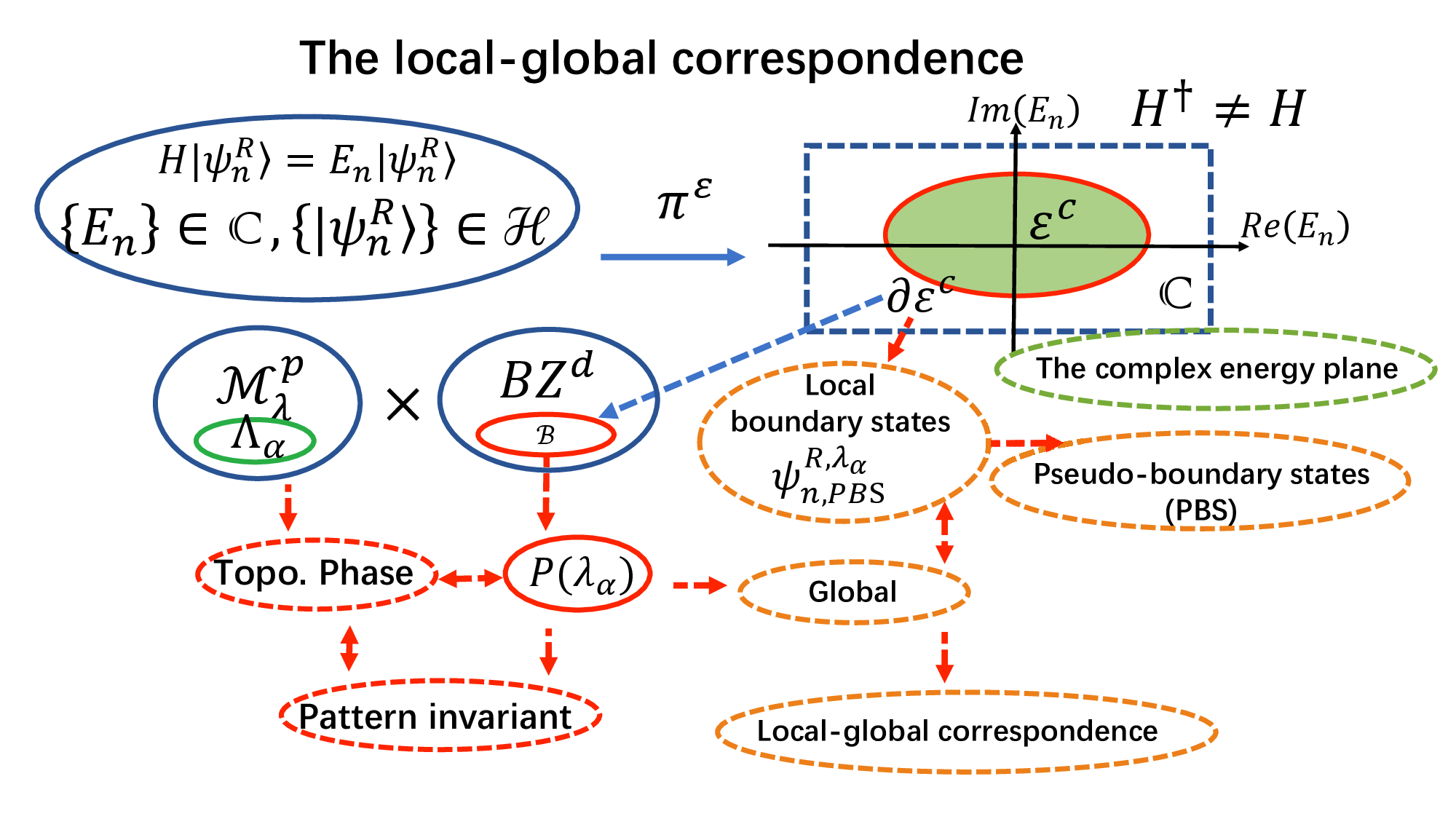}
	\caption{Online color: The schematic representation of the local-global correspondence between the PBS in the complex energy plane and the topological invariants of  non-Hermitian systems. For a given set of parameters $\lambda\in \mathcal{M}^{p}_{\lambda}$, the eigen energies of the non-Hermitian Hamiltonian can be transformed into the complex energy plane, $\mathcal{E}^{c}\subset \mathbb{C}$, to form a band (green region), in which the red curve $\partial\mathcal{E}^{c}$ denotes the boundary states of the bulk band. We refer to these boundary states located at $\partial\mathcal{\varepsilon}^c$ as the PBS.}
	\label{fig1}
\end{figure}

Let us first set up a conceptual framework to present a local-global correspondence between the boundary states in the complex energy plane and the topological invariants of  the non-Hermitian system. The schematic illustration is given in Fig.\ref{fig1}.
\begin{itemize}
  \item For given $(\mathbf{k},\lambda)\in BZ^{d}\times \mathcal{M}_{\lambda}^{p}$,
  the eigen equation of the Hamiltonian gives the eigen energy bands and their corresponding eigen vectors. The complex eigen energy bands can be mapped according to $\pi^{\varepsilon}: E_{n}(\mathbf{k},\lambda)\rightarrow\left(E_{n,R}(\mathbf{k},\lambda), E_{n,I}(\mathbf{k},\lambda)\right)\in\mathcal{E}^{c}$, where $\mathcal{E}^{c}$ is the complex energy plane and $\left(E_{n,R}(\mathbf{k},\lambda), E_{n,I}(\mathbf{k},\lambda)\right)$ are the real and imaginary parts of the energy band respectively.

  \item For given a specific parameter $\lambda\in \mathcal{M}_{\lambda}^{p}$, the complex energy band in the BZ is mapped to the complex energy plane in general forms a few of the two-dimensional regions in the complex energy plane, labeled by $\mathcal{E}^c$, which are called the bulk bands.\cite{Fu}

  \item The phase diagram of the non-Hermitian system can be represented by a union of subspaces in the parameter space,
   $\Lambda=\bigcup_{\alpha}\Lambda_{\alpha}$, where $\lambda_{\alpha}\in\Lambda_\alpha$ denotes the topological phase $\alpha$ which are characterized by a topological index such as the winding number $w_{\alpha}$, Chern number $C_{\alpha}$ or vorticity $\nu_{\alpha}$.\cite{Ghatak}

   \item For a given energy band we define some states $\psi^{R,\lambda_{\alpha}}_{n,PBS}$ of the bulk band, whose
   their corresponding eigen energies are located at the boundary of the energy band $\mathcal{E}^c$,

     \begin{equation}\label{EDS1}
     \psi^{R,\lambda_{\alpha}}_{n,PBS} := \left\{\psi^{R}_{n}(\mathbf{k},\lambda) \left|
     \begin{array}{l}
      H\psi^{R}_{n}(\mathbf{k},\lambda)=E_{n}\psi^{R}_{n}(\mathbf{k},\lambda), \\
      E_{n}(\mathbf{k},\lambda)=(E_{n,R},E_{n,I})\in \partial \mathcal{E}^{c} \\
      \mathbf{k}\in \mathcal{B}\subset BZ, \\
      \lambda\in \Lambda_{\alpha}\subset \mathcal{M}^{p}_{\lambda},
     \end{array}\right.
     \right\}
     \end{equation}

      In other words, the subset in the BZ $\mathcal{B}$ is $\mathcal{B}\leftarrow\left\{\psi^{R,\lambda_{\alpha}}_{n,PBS}\right\}\leftrightarrow (E_{n,R},E_{n,I})\in\partial \mathcal{E}^{c}\subset\mathcal{E}^{c}$, where $\partial \mathcal{E}^{c}$ is the boundary of the bulk band. We refer to $\partial \mathcal{E}^{c}$ ($\psi^{R,\lambda_{\alpha}}_{n,PBS}$) as the pseudo boundary state (PBS) in the complex energy plane.
      In general, the bands $\mathcal{E}^{c}$ vary with the parameters, leading to variation of the PBS.

 \item For a given topological phase in the parameter space, $\Lambda_{\alpha}$, the PBS $\left\{\psi^{R,\lambda_{\alpha}}_{n,PBS}\right\}$ mapped to the BZ form a pattern, which is defined as
     \begin{equation}\label{PP0}
     P(\lambda) := \left\{(k_x,k_y)\left|
     \begin{array}{l}
     \mathbf{k}\in \mathcal{B}\subset BZ,\\
     \lambda\in \Lambda_{\alpha}\subset \mathcal{M}^{p}_{\lambda},\\
     \mathcal{B}\leftarrow \left\{\psi^{R,\lambda_\alpha}_{n,PBS}\right\},
     \end{array}\right.
     \right\}.
     \end{equation}
 $P(\lambda)$ and $\mathcal{B}$ are in one-to-one correspondence with each other. $P(\lambda)$ denotes the pattern of the PBS in the BZ, $\mathcal{B}$ is the position of the PBS in the BZ, and the PBS is the set of boundary states in the complex energy plane. Their concrete representations can be seen in the sub figures (j) to (m) of Fig.(\ref{fig2}).

\end{itemize}

It should be remarked that the PBS as given in (\ref{EDS1}) is well-defined when the wave vector in the BZ is continuous. In practice, we obtain the PBS using the numerical methods (see Appendix A), in which the approximation depends on the discrete precision of the BZ and the input of the radius of the ball pivoting algorithm.\cite{Fausto} However, the approximation of the PBS only affects the precision of the pattern of PBS in the BZ, and does not affect the structure of the pattern when the parameters are within the gapped regions. In other words, the numerical method and its approximation does not change the results because the topological invariant is robust and the numerical method is efficient.

\subsection{The local-global correspondence}

Based on above conceptual and mathematical frameworks, we propose a local-global correspondence between the PBS and topological invariants.

\textbf{The local-global correspondence: }
Consider a generalized non-Hermitian Dirac model with the Hamiltonian, $H(\mathbf{k},\lambda)=\mathbf{h}(\mathbf{k},\lambda)\cdot\sigma$, where $\mathbf{h}(\mathbf{k},\lambda)\in \mathbb{C}$. Suppose that there exist two kinds of energy band structures in the parameter space, gapped and gapless, which denote $\Lambda_\alpha$ and $\Lambda_\beta$, respectively.

\begin{itemize}
  \item The PBS $\left\{\psi^{R,\lambda_{\alpha}}_{n,PBS}\right\}$ of the energy band in the complex energy plane determines the topological invariants in the parameter space. The pattern $P(\lambda_{\alpha})$ in the BZ is a topological invariant in the gapped phases,
 \begin{subequations} \label{PP1}
   \begin{eqnarray}
   P(\lambda'_{\alpha})&=&P(\lambda_{\alpha}) \quad \textrm{for} \quad\forall \lambda_{\alpha},\lambda'_{\alpha}\in \Lambda_{\alpha}\\
      \mathcal{B}&\leftarrow & \left\{\psi^{R,\lambda_\alpha}_{n,PBS}\right\}\leftrightarrow\partial \mathcal{E}^{c},
   \end{eqnarray}
 \end{subequations}
 where $\Lambda_\alpha$ denotes the gapped phases.
\item The $P(\lambda_{\beta})$ of the PBS in the BZ vary in the gapless phases,
 \begin{subequations} \label{PP2}
   \begin{eqnarray}
   P(\lambda'_{\beta})&\neq &P(\lambda_{\beta}) \quad \textrm{for} \quad\forall \lambda_{\beta},\lambda'_{\beta}\in \Lambda_{\beta}\\
      \mathcal{B}&\leftarrow & \left\{\psi^{R,\lambda_\beta}_{n,PBS}\right\}\leftrightarrow\partial \mathcal{E}^{c},
   \end{eqnarray}
 \end{subequations}
 where $\Lambda_\beta$ denotes the gapless phases.
     The boundary of the topological phase transition can be detected by changes is the pattern $P(\lambda)$ when the parameters vary, crossing the boundary between different topological phases.

\end{itemize}

It should be remarked that (1) the PBS is composed of only some of the eigen states of the energy band in the complex energy plane, but that dominates the topological invariants against the parameter deformation in the region of the topological phase. This implies that there exists a correspondence between the PBS (local information) and the topological invariant (global property of systems). Thus, we refer to this correspondence between the PBS and the topological invariant defined in ({\ref{PP1}}) as the local-global correspondence, which can be regarded as a generalized bulk-boundary correspondence in the complex energy plane. (2) In the gapless phase, two bulk bands in the complex energy plane combine together such that some of the PBS disappear and become bulk states.  The pattern $P(\lambda)$ of the PES in the BZ becomes unstable as the parameters vary. (3) We found that the length and the analytic behavior of the PBS change discontinuously at the boundary between the gapped and gapless phase, which provides a criterion for the topological phase transition. The detail description was presented elsewhere.\cite{Annan1} (4) For Hermitian cases, the energy bands are real and the complex energy plane disappears. Thus, this approach is not available for Hermitian cases.

\subsection{The homotopic continuous maps}
\textbf{The homotopic continuous maps:} For given $(\mathbf{k},\lambda)\in BZ^{d}\times \mathcal{M}^{p}_\lambda$, we define two continuous maps,
\begin{subequations} \label{hg}
\begin{eqnarray}
f(\lambda)&:& \partial \varepsilon^{c}(\lambda) \rightarrow  P(\lambda) \sim \mathcal{B}\subset BZ\\
g(\lambda')&:& \partial \varepsilon^{c}(\lambda') \rightarrow  P(\lambda')\sim \mathcal{B}\subset BZ
\end{eqnarray}
\end{subequations}
where $\forall\lambda,\lambda'\in \Lambda_\alpha$. In general,
These two maps produce two patterns of the PBS in the BZ. The patters vary with varying the parameters.
On the other hand, we define two homotopic functions,\cite{Eber}

\begin{equation}\label{HH1}
H(\lambda,\xi)=(1-\xi)f(\lambda)+\xi g(\lambda), \quad\forall\lambda \in \Lambda_\alpha ,
\end{equation}
where $\xi\in [0,1]$. When $\xi=0$, $H(\lambda,\lambda)=f(\lambda)$ and $\xi$ varies from $0$ to $1$, $H(\lambda,\xi)$ varies from $f(\lambda)$ to $g(\lambda)$ continuously. For $\forall\lambda,\lambda'\in \Lambda_\alpha$ we have $P(\lambda)=P(\lambda')$, namely $f(\lambda)$ and $g(\lambda)$ are homotopic continuous maps and the pattern $P(\lambda)$ is a topological invariant under homeomorphisms. However, for $\forall\lambda\in \Lambda_\alpha$ and $\forall\lambda'\in \Lambda_\beta$, where $\Lambda_\alpha\cap \Lambda_\beta=\emptyset$, we have $P(\lambda)\neq P(\lambda')$, namely $f(\lambda)$ and $g(\lambda)$ are not homotopic continuous maps.
This mathematical representation of the continuous map tells us that the topological invariant of quantum phase in the parameter space is a homotopic equivalence.

The detailed description of the local-global correspondence depends on the concrete model. We will demonstrate this correspondence using a typical non-Hermitian model, Chern insulator model, in the following section. In general, the energy band near the Fermi energy play a crucial role in dominating the physical properties of the system. The local-global correspondence reveals that only some of the states in the energy band near the Fermi energy dominate the topological invariants and their phase transitions.

\section{Non-Hermitian Chern insulator model}
\subsection{Topological phase diagram}
To test the local-global correspondence, we will repeat the phase diagram of the non-Hermitian Chern insulator model to demonstrate the validity of the local-global correspondence. The non-Hermitian Chern insulator model contains rich physics and has been studied extensively. \cite{Kohei2,Ghatak} Let us first recall some basic properties of the non-Hermitian Chern insulator model. The Bloch Hamiltonian of the non-Hermitian Chern insulator is given by\cite{Kohei2,Ghatak}
\begin{equation}\label{CS1}
H(\mathbf{k})=\mathbf{h}(\mathbf{k})\cdot \mathbf{\sigma}
\end{equation}
where 
\begin{subequations}\label{hhh1}
\begin{eqnarray}
h_{x}(\mathbf{k}) &=& t\sin k_{x},\\
h_{y}(\mathbf{k}) &=& t\sin k_{y}-i\delta \\
h_{z}(\mathbf{k}) &=& m+t\cos k_{x}+t\cos k_{y},
\end{eqnarray}
\end{subequations}
where $t,m,\delta\in \mathcal{M}^{3}_{\lambda}$ are real parameters and $t\in \mathbb{R}^+$. $(k_x,k_y)\in [-\pi,\pi]^2$. The energy bands of the model are obtained

\begin{equation}\label{EECS}
E_{\pm}=\pm\sqrt{m^2-\delta^2+2t^2(1+\cos k_{x}\cos k_{y})+2mt(\cos k_{x}+\cos k_{y})-2it\delta\sin k_{y}}.
\end{equation}

The real and imaginary parts of the energy bands are obtained by
\begin{subequations}\label{RIEB1}
\begin{eqnarray}
E_{R} &=& \sqrt{\frac{\epsilon+\sqrt{\epsilon^2+\omega^2}}{2}},\\
E_{I} &=& \sqrt{\frac{-\epsilon+\sqrt{\epsilon^2+\omega^2}}{2}},
\end{eqnarray}
\end{subequations}
where
\begin{subequations}\label{RIEB2}
\begin{eqnarray}
\epsilon &=& 2t^2+m^2-\delta^2+2t^{2}\cos k_x\cos k_y+2mt(\cos k_x+\cos k_y),\\
\omega &=& -2t\delta \sin k_y.
\end{eqnarray}
\end{subequations}
The complex energy bands can be rewritten as
\begin{equation}\label{CEB2}
E_{\pm}=\pm (E_R+\sigma i E_I),
\end{equation}
where $\sigma=1$ for $\omega\geq 0$ and $\sigma=-1$ for $\omega < 0$.

The eigen vectors of the Hamiltonian for the non-Hermitian Chern insulator model are given by\cite{Kohei2,Ghatak}
\begin{equation}\label{EV1}
|\psi^{R}_{\pm}\rangle=\frac{1}{\sqrt{2E_{\pm}(E_{\pm}-h_{z})}}\left(
\begin{array}{c}
h_{x}-ih_{y}\\
E_{\pm}-h_{z}
\end{array}
\right)
\end{equation}
and its dual vectors are
\begin{equation}\label{EV2}
\langle \psi^{L}_{\pm}|=\frac{1}{\sqrt{2E_{\pm}(E_{\pm}-h_{z})}}\left(
\begin{array}{cc}
h_{x}+ih_{y} & E_{\pm}-h_{z}
\end{array}
\right).
\end{equation}

The topological phase transition occurs when the complex energy band closes, such that $E_{R}=E_{I}=0$. Using (\ref{RIEB1})-(\ref{RIEB2}), the quantum phase transitions happen at the exceptional points, $k_{y}=0,\pm\pi$. The phase transition boundary is given by \cite{Kohei2}
\begin{subequations}\label{BC1}
\begin{eqnarray}
|m|\leq |\delta| \leq |m+2t| \quad \textrm{for} \quad m\geq -t \\
|m+2t|\leq |\delta| \leq |m| \quad \textrm{for} \quad m\leq -t,
\end{eqnarray}
\end{subequations}
for $k_{y}=0$ and
\begin{subequations}\label{BC2}
\begin{eqnarray}
|m-2t| \leq |\delta| \leq  |m|\quad \textrm{for} \quad m\geq t \\
|m|\leq |\delta| \leq |m-2t| \quad \textrm{for} \quad m\leq t,
\end{eqnarray}
\end{subequations}
for $k_{y}=\pm\pi$. Let us suppose that $t=1$ for convenience without loss of generality. The exceptional lines (\ref{BC1}) and (\ref{BC2}) separate different kinds of quantum phases, such as gapped and gapless phases in the $m-\delta$ plane. The gapped phase is characterized by the Chern number, \cite{Kohei2,Ghatak}
\begin{equation}\label{CN1}
C_{\pm}= \frac{1}{2\pi}\int_{\textrm{BZ}}\Omega^{\pm}(\mathbf{k})d^{2}\mathbf{k}
\end{equation}
where $\Omega^{\pm}(\mathbf{k})$ is the Berry curvature, which is given by \cite{Annan}
\begin{equation}\label{BC3}
\Omega^{\pm}(\mathbf{k}) = \mp\frac{1}{2}\mathbf{\widehat{h}}\cdot
\left(\frac{\partial \mathbf{\widehat{h}}}{\partial k_x}\times\frac{\partial \mathbf{\widehat{h}}}{\partial k_y}\right)
\end{equation}
where $\mathbf{\widehat{h}}$ is the unit vector of $\mathbf{h}$.

Interestingly, for non-Hermitian systems, the Chern number could be complex in general. The quantum Hall conductance can be generalized to quantum hall admittance for the non-Hermitian Dirac model.\cite{Annan} The imaginary part of the Chern number can be interpreted as susceptance, which implies that intrinsic capacitance and induction could emerge in non-Hermitian systems. However, for the non-Hermitian Chern insulator model, numerical results indicate that the imaginary part of the Chern number is zero in the gapped phase.\cite{Kohei2,Ghatak,Annan2}

\begin{figure}[htbp]
	\centering
	\includegraphics[width=1\linewidth]{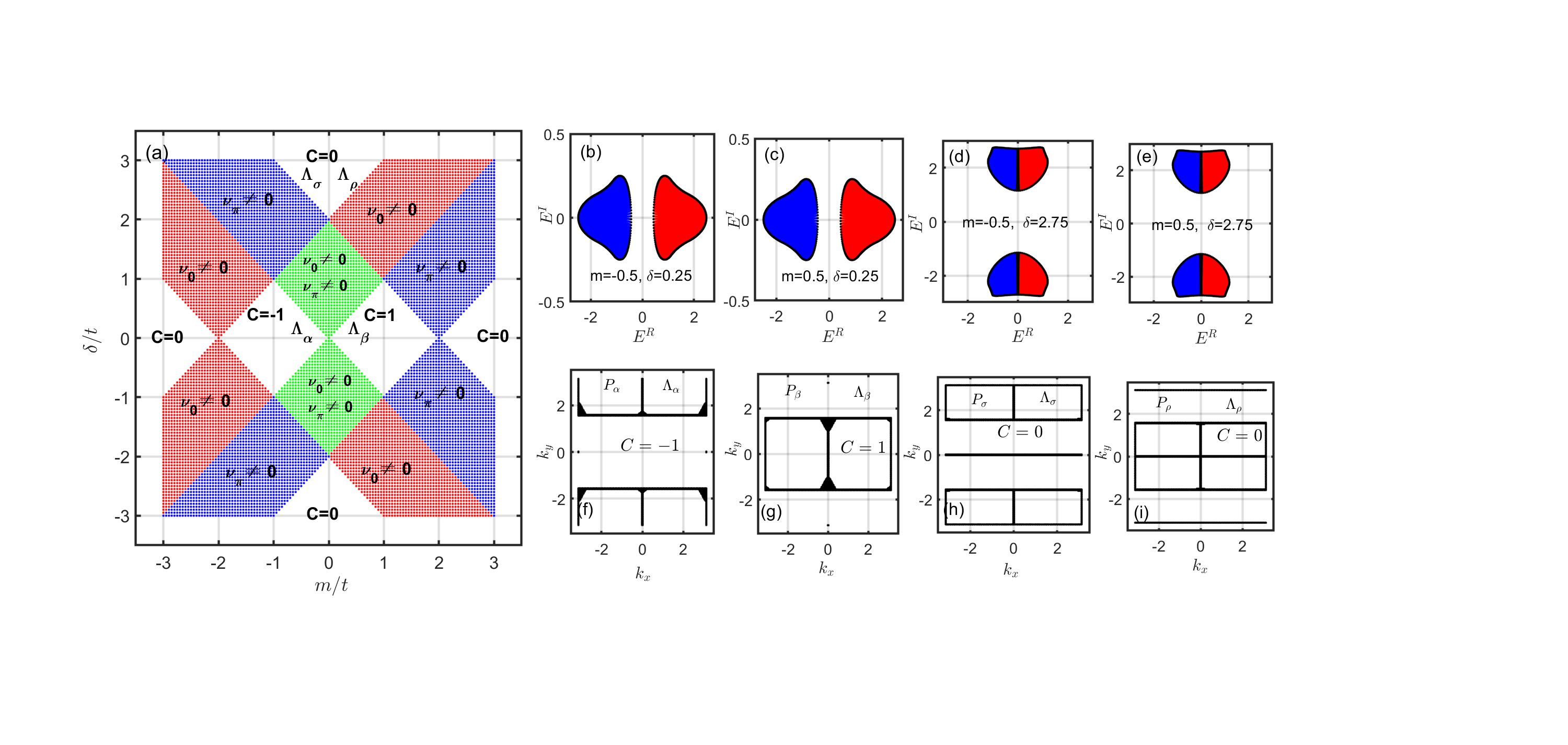}
	\caption{ Online Color: (a) The topological phase diagrams, in which $C=0,\pm 1$ and $\nu_{0(\pi)}\neq 0$ characterize the topological gapped and gapless phases with the white (gapped) and color (gapless) regions, respectively.\cite{Kohei2} (b)-(e) The complex energy bands are transformed to the complex energy plane for the parameters, $m=\pm 0.5, \delta=0.25,2.75$, respectively, which correspond to the gapped phases $C=\pm 1, 0$.
 (f)-(i) The patterns of the PBS in the BZ for the parameters in the ranges $\Lambda_\alpha,\Lambda_\beta,\Lambda_\sigma,$ and $\Lambda_\rho$. The flaws in the line are numerical approximations .(see Appendix A) }
	\label{fig2}
\end{figure}

The gapless phase can be characterized by the vorticity, which is defined by \cite{Kohei2,Ghatak}
\begin{equation}\label{VT1}
\nu_{mn}(\mathbf{k}_{\textrm{EP}})=\frac{1}{2\pi }\oint_{\mathcal{C} (\mathbf{k}_{\textrm{EP}})}\nabla_{\mathbf{k}} \arg [E_m(\mathbf{k})-E_n(\mathbf{k})] \cdot d\mathbf{k}
\end{equation}
where $\mathcal{C} (\mathbf{k}_{\textrm{EP}})$ in (\ref{VT1}) denotes a loop encircling the exceptional point in the BZ.
Fig.\ref{fig2}(a) shows the phase diagram obtained from the exceptional lines in (\ref{BC1}) and (\ref{BC2}),\cite{Kohei2,Ghatak} in which the white regions are the topological gapped phase characterized by the Chern number $C=0, \pm 1$ and the other colored regions are the gapless phases characterized by the vorticity $\nu$. \cite{Kohei2,Ghatak}
The phase diagrams of the Chern insulator model in the parameter space based on the exceptional lines in (\ref{BC1}) and (\ref{BC2}) are shown in Fig.\ref{fig2}(a).

\subsection{The local-global correspondence and topological invariants}
The local-global correspondence in the complex energy plane provides a novel way to explore the topological invariants of non-Hermitian systems.
This approach is implemented based on the conceptual and mathematical framework presented in the Sec. II. The detailed numerical algorithm is presented in the Appendix A.
The complex energy bands are transformed to the complex energy plane in Figs.\ref{fig2}(b)-(e), in which the blue and red parts of the complex energy bands are negative and positive, respectively. The patterns of the PBS in the BZ are plotted in Figs.(\ref{fig2}) (f)-(i), which correspond to the boundary of the complex energy bands in Figs.\ref{fig2}(b)-(e).
We find that the patterns are robust and topological invariant in the gapped topological phases, $\Lambda_\alpha, \Lambda_\beta, \Lambda_\sigma$, and $\Lambda_\rho$.The patterns in the gapless phase vary with the varying parameters. At the boundary between the gapped and gapless phases, the patterns become unstable and sensitively depend on the parameter values.\cite{Annan2}
It should be pointed out that when we investigate the patterns in the BZ we should note the translation symmetry and the periodicity of the BZ.
Thus, the patterns in Figs.\ref{fig2}(h) and (i) are topologically equivalent, and have the same lengths. They become exactly the same due to the translation symmetry of the BZ. Namely, when the BZ is translated from the $\Gamma$ to $M$ point, the patterns in Figs.\ref{fig2}(h) and (i) become the same. \cite{Annan2} Similarly the patterns in Figs.\ref{fig2}(f) and (g) are also topologically equivalent. These results are consistent with the previous results obtained using the Chern number and winding number. \cite{Kohei2,Ghatak}

\begin{figure}[htbp]
	\centering
	\includegraphics[width=1\linewidth]{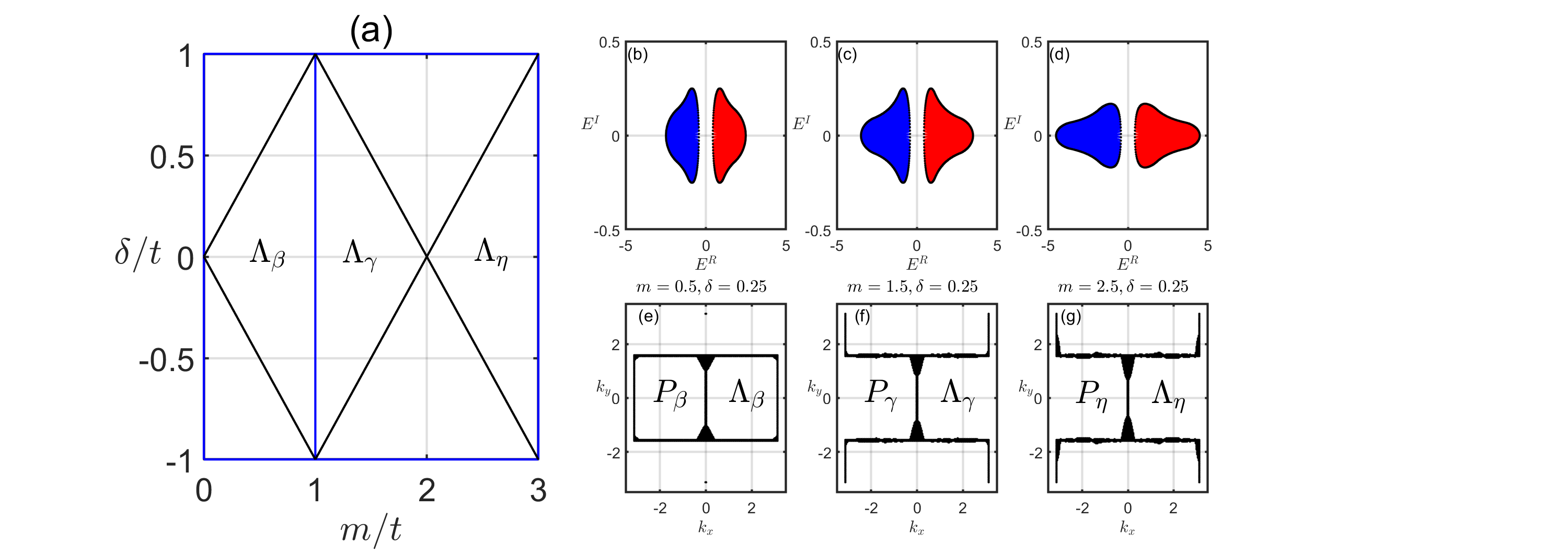}
	\caption{Online color: (a) The sub-phase diagram in the parameter space within $0\leq m/t\leq 3$ and $-1\leq \delta/t\leq 1$. (b)-(d) The complex energy bands are transformed to the complex energy band plane for the parameters, $m= 0.5,1.5$ and $2.5$ with  $\delta=0.25$, respectively. (e)-(g) The patterns of the PBS in the BZ for the corresponding parameter regions, $\Lambda_\beta, \Lambda_\gamma$ and $\Lambda_\eta$, in which the flaws on the line are numerical errors (see Appendix A)}
	\label{fig3}
\end{figure}

Interestingly, when we investigate the topological invariants of the patterns of PBS in the BZ in the parameter regions $m= 0.5,1.5$ and $2.5$, with $\delta=0.25$, in Fig.\ref{fig3}, we find a novel topological invariant embedded in these regions beyond that given by the Chern number.
In Fig.\ref{fig3}(a) we plot a part of the topological phase diagram to compare the topological invariants using both of the local-global correspondence and the Chern number.
We plot the energy bands in the complex energy plane and the patterns of the PBS in the BZ for different parameter regions in Fig.\ref{fig3}(b)-(d), in which the regions of the three bands looks similar, but their PBS are quite different. The patterns of the PBS are plotted in Figs.\ref{fig3} (e)-(g). We find that the patterns in Fig.\ref{fig3} (f) and (g) are exactly the same and are robust in the parameter region, $\Lambda_\gamma$ and $\Lambda_\eta$. This implies that the quantum states in the parameter regions of $\Lambda_\gamma$ and $\Lambda_\eta$ contain some of the same topological invariants, but that these are different from one in the parameter region $\Lambda_\beta$. All of these properties  are the same in the negative $m$ region in Fig.\ref{fig3}.

For the gapless region in the phase diagram, $\nu_{0(\pi)}\neq 0$, the patterns of the PBS in the BZ vary with the varying parameters. Hence, the energy band gap plays an essential role in protecting the topological phases of non-Hermitian systems. At the boundary between the gapped and gapless regions, the patterns of the PBS are sensitively unstable and depend on the variation of parameters. These properties of the patterns of the PBS provide a signal which allows us to detect the topological phases of non-Hermitian systems.

It should be pointed out that the topological phases in $\Lambda_\beta$ and $\Lambda_\gamma$ were characterized by the Chern number $C=1$, but $C=0$ in $\Lambda_\eta$. This implies that the topological invariants of the quantum states in $\Lambda_\beta$ and $\Lambda_\gamma$ are the same, but different from $\Lambda_\eta$.
The topological invariants based on the local-global correspondence enrich the topological phase diagram of the non-Hermitian Chern insulator model. The quantum states in the regions $\Lambda_\beta$ and $\Lambda_\gamma$ have the same topological invariants labeled by the Chern number $C=1$, which is different from that of the region $\Lambda_\eta$, labeled by the Chern number $C=0$.\cite{Kohei2,Ghatak} However, We found from the local-global correspondence that the patterns of the PBS in the regions $\Lambda_\gamma$ and $\Lambda_\eta$ are exactly same and robust even though their Chern numbers are different. This reveals a novel topological invariant in the regions $\Lambda_\gamma$ and $\Lambda_\eta$ beyond that predicted by the Chern number.\cite{Kohei2,Ghatak} We will further demonstrate this point using the generalized vorticity method in the following subsection. From a mathematical point of view, there exist different topological structures in the same manifold, which can be detected by different methods.

\subsection{Generalized vorticity and topological invariants}
\begin{figure}[htbp]
	\centering
	\includegraphics[width=1\linewidth]{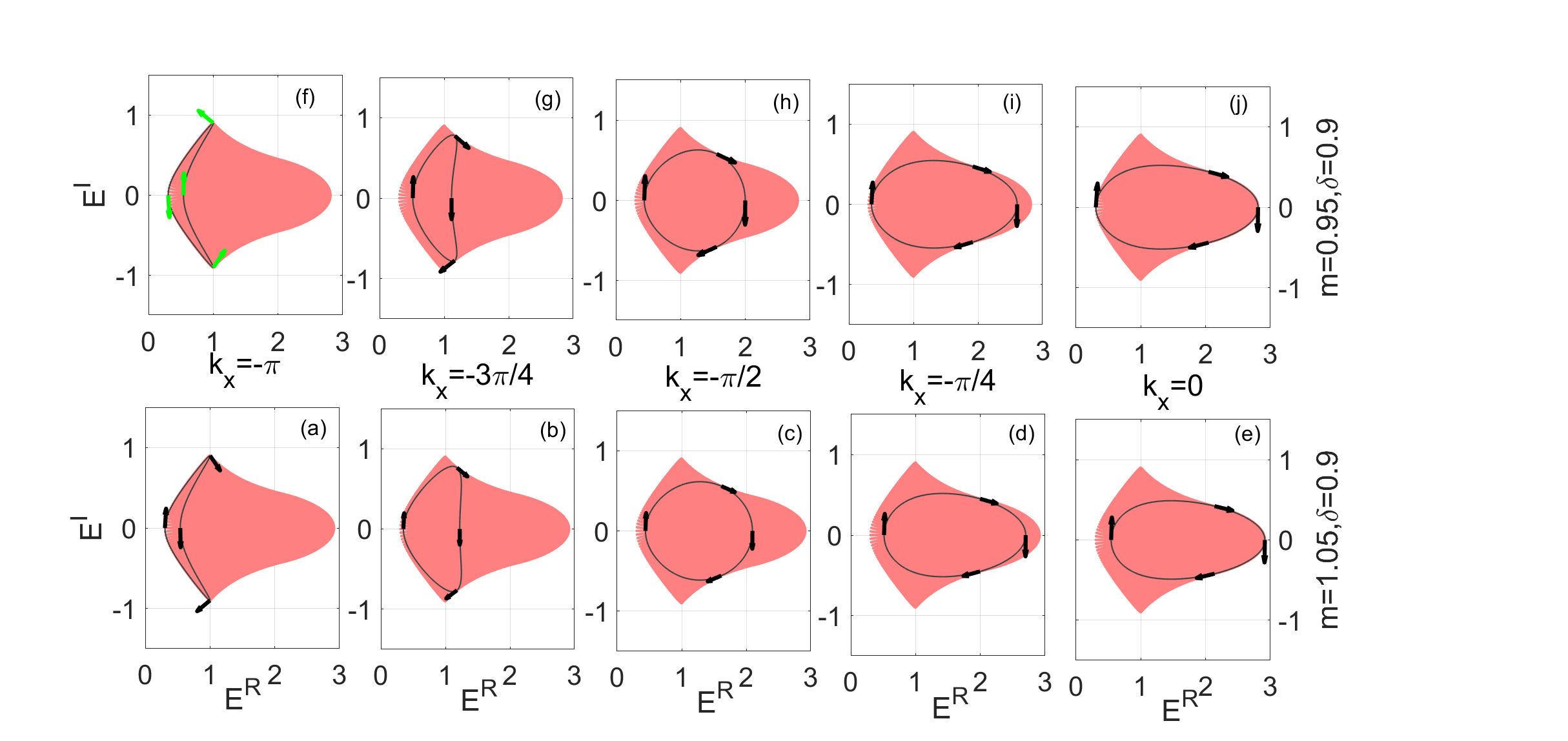}
	\caption{Online color: The generalized vorticity $\mu_{n}(k_x)$ in the complex energy plane from $k_x=-\pi$ to $0$. In (a)-(e) the $\mu_{n}(k_x)$ are for $m=1.05$ and $\delta=0.9$ and in (f)-(j) the $\mu_{n}(k_x)$ are for $m=0.95$ and $\delta=0.9$. It should be noted that the arrows in (a)-(e) are clockwise, but the arrow (green) in (f) is anticlockwise, which implies that the loop of the $\mu_{n}(k_x)$ for $m=0.95$ flips at some point between $k_x=-\pi$ and $-3\pi/4$.}
	\label{fig4}
\end{figure}

The local-global correspondence reveals that the pattern of the PBS is topological invariant against parameter deformations in the topological phase.
To understand the physical mechanism of the PBS behind the local-global correspondence, we introduce a generalized $k_x$-dependent vorticity for given $k_x$ in the BZ

\begin{equation}\label{UN1}
\mu_{n}(k_x) :=\frac{1}{2\pi}\int_{-\pi}^{\pi}\partial_{k_y}\vartheta_{n}(k_x,k_y) dk_y
\end{equation}
where
\begin{equation}\label{EE2}
 \vartheta_{n}(k_x,k_y)=\arctan\frac{\Im\left(E_{n}(k_x,k_y)-\frac{1}{2}\left[E_{n}(k_x,0)+E_{n}(k_x,\pi)\right]\right)}
 {\Re\left(E_{n}(k_x,k_y)-\frac{1}{2}\left[E_{n}(k_x,0)+E_{n}(k_x,\pi)\right]\right)}.
\end{equation}
This generalized $k_x$-dependent vorticity describes a $k_y$-directional circle in the complex energy plane.
Fig. (\ref{fig4}) shows the loop of the generalized $k_x$-dependent vorticity in the complex energy plane from $k_x=-\pi$ to $0$ for the parameters $m=1.05, 0.95$ and $\delta=0.9$. Here we plot only a half of the range of $k_x$ due to the parity symmetry of $\pm k_x$ for the $\mu_{n}(k_x)$.
The arrows of the loops are clockwise from the insets (a) to (e) in Fig.(\ref{fig4}) for $m=1.05$, but for $m=0.95$ the arrow (green) in the inset (f) is anticlockwise, which implies the loop flips at some point between $k_x=-\pi$ and $-3\pi/4$ for the parameter $m=0.95$. The flipping point $k_x$ depends on the parameters $m$ and $\delta$, but the number of flippings is robust against parameter deformations, and can be viewed as a topological index.
Thus, we define the flipping index of this generalized $k_x$-dependent vorticity,

\begin{subequations}\label{GV1}
\begin{eqnarray}
\Theta^R &=& \left|\Re [\mu(0)-\mu(\pi)]\right|\\
\Theta^I &=& \left|\Im [\mu(0)-\mu(\pi)]\right|
\end{eqnarray}
\end{subequations}
to count the flipping number of the loop in the complex energy plane within the regions $-\pi\leq k_x\leq 0$ or $0\leq k_x\leq \pi$. The flipping index is an integer, which can be regarded a counterpart of Chern number induced by the edge states crossing at the Fermi energy in Hermitian systems. The numerical calculations show the $\Theta^{R(I)}$ in the parameter space. It can be seen that the real flipping index is $\Theta^R =1$ in the region $\Lambda_\beta$ and whereas $\Theta^R =0$ in the regions $\Lambda_\gamma$ and $\Lambda_\eta$. These results are consistent with the results obtained by the local-global correspondence. The real flipping index is $\Theta^R =1/2$ in the regions of the upper and lower triangles, which corresponds to the Chern number $C=0$. The imaginary flipping index is $\Theta^I =0$ in the gapped region $\Lambda_\gamma$ and $\Lambda_\eta$.

\begin{figure}[htbp]
	\centering
	\includegraphics[width=1\linewidth]{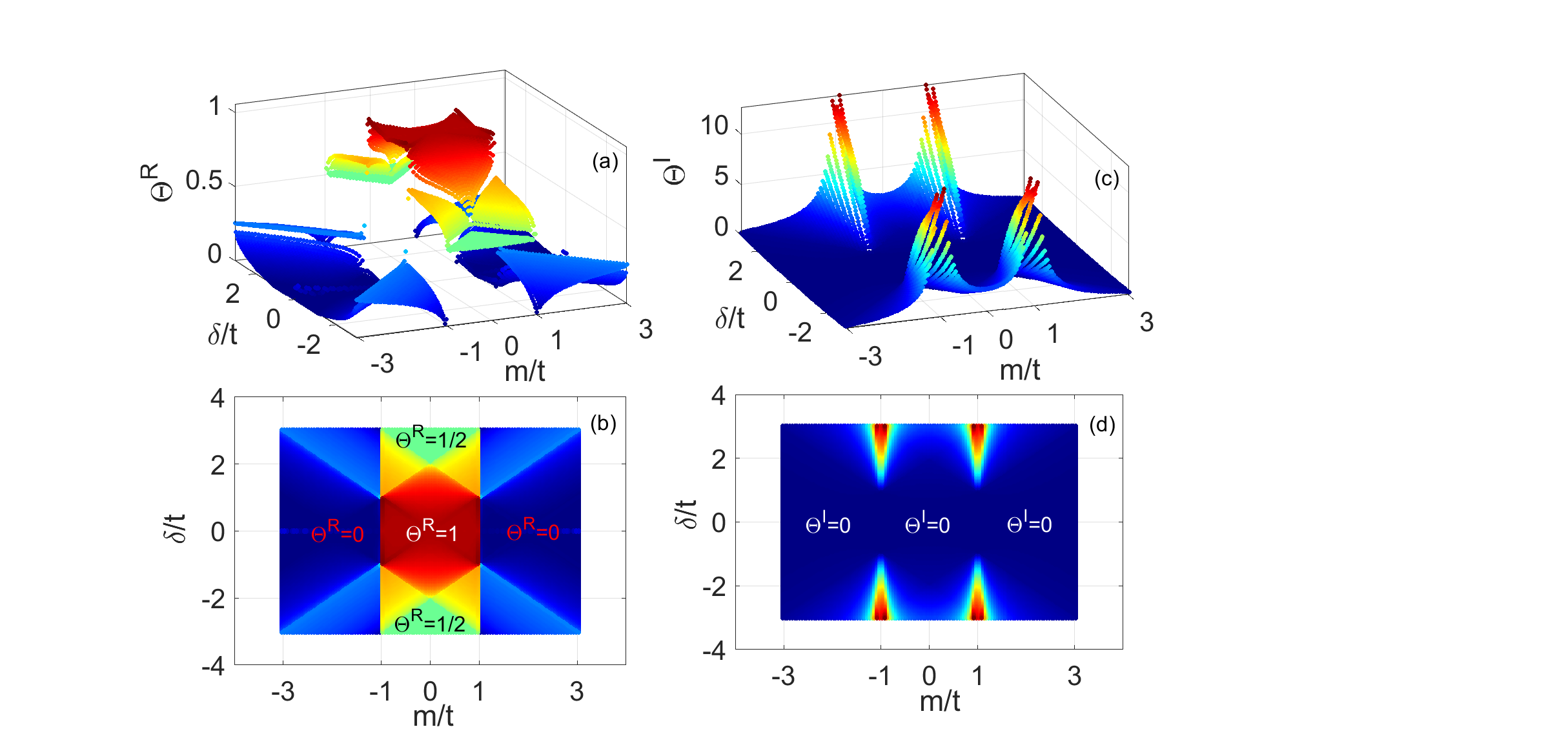}
	\caption{Online color: The flipping index of the generalized vorticity $\Theta$ in the parameter space. (a) The real part of $\Theta$ and (b) its projection (phase diagram) in the parameter space. (c) The imaginary part of $\Theta$ and its projection (phase diagram) in the parameter space. }
	\label{fig5}
\end{figure}

It should be remarked that the flipping index of the generalized vorticity is compatible with the vorticity of non-Hermitian systems.
In principle, we can also define the generalized $k_{y}$-dependent vorticity $\mu_{n}(k_y)$. However, the wave vector-dependent vorticities $\mu_{n}(k_x)$ and $\mu_{n}(k_y)$
are not always equivalent. In practice, the choices of the wave vector directions depends on the concrete model. Moreover, it is worth studying what physical observables emerge from the varying of this generalized vorticity, in particulars, what relationship between the variation of the local flipping point and the global topological invariant of the flipping index exists and what physical phenomena emerge from the analytic behaviors of the imaginary part of the generalized vorticity.

\section{Remarks on topological invariants}
\subsection{Relationship between the Chern number and the local-global correspondence}

Mathematically, the existence of a topological invariant means that some objects (here quantum states or phases) are
topological (or homotopy) invariant under homeomorphisms or are the same for the homotopically equivalent topological space, where the homeomorphism and homotophically equivalent classes are some regions in the parameter space.

A quantum system can be described by some topological quantum numbers, such as the Berry phase, Chern number, winding number and vorticity, which characterize
different equivalence classes of quantum states in the parameter space.
The Chern number depends on the exceptional points, which correspond to the closure of the energy band. The energy band gap protects the topological invariance of the quantum states.
The local-global correspondence depends on the PBS in the complex energy plane, which can be seen from the generalized vorticity and its corresponding flipping index of the wave-vector-dependent vorticity. There are two physical mechanisms that connect the Chern number and the local-global correspondence. One is the complex phase of the complex energy band, changes in which are induced by the generalized vorticity and its corresponding flipping number. This can depend on the energy band crossing instead of closing. The other is the relationship between the closure of energy bands and the changes of the PBS. When the energy bands close, their corresponding regions in the complex energy plane change from two separate regions to one connected region, which can lead to the PBS and their corresponding patterns in the BZ changing. These two mechanisms are independent and give rise to different parameter-dependent topological invariants of the quantum states.

\subsection{Transformation between the BZ and the complex energy plane}

The local-global correspondence involves the transformation between the BZ and the complex energy plane.
Let us now investigate this transformation in greater detail.

The differential area mapped from $BZ^{2}$ to $\mathcal{E}^{c}$ for a given set of parameters is represented as a transformation from the BZ to the complex energy plane,
\begin{equation}\label{TM1}
\left(
\begin{array}{c}
dE_{R} \\
dE_{I}
\end{array}
\right)=
J(\mathbf{k},\lambda)
\left(
\begin{array}{c}
dk_x \\
dk_y
\end{array}
\right),
\end{equation}
where $J(\mathbf{k},\lambda)$ is the transformation matrix, which is given by
\begin{equation}\label{JD1}
J(\mathbf{k},\lambda)=\left(
\begin{array}{cc}
\frac{\partial E_{R}}{\partial k_{x}} & \frac{\partial E_{R}}{\partial k_{y}} \\
\frac{\partial E_{I}}{\partial k_{x}} & \frac{\partial E_{I}}{\partial k_{y}}
\end{array} \right)
\end{equation}
The Jacobian determinant of the transformation matrix $\det J(\mathbf{k},\lambda)$ is either nonzero or zero depending on whether the transformation is invertible or not invertible, respectively. In other words, when the Jacobian determinant is zero, the transformation is not one-to-one.

For the non-Hermitian Chern insulator model, note that for the relationship between the real and imaginary parts of the energy band in (\ref{RIEB1}) and (\ref{RIEB2}), we have
$E^{2}_{R}-E^{2}_{I}=\epsilon$ and $E^{2}_{R}E^{2}_{I}=\omega^{2}/4$ , yielding

\begin{subequations}\label{ERI1}
\begin{align}
E_{R}^{2}-E_{I}^2 &= 2-\delta^{2}+m^{2}+2\cos k_{x}\cos k_{y}+2m(\cos k_{x}+\cos k_{y}), \\
E_{R}E_{I} &= \delta\sin{k_{y}}.
\end{align}
\end{subequations}
where we set $t=1$ for convenience. Taking derivative with respect to $k_x$ and $k_y$ in (\ref{ERI1}),
for the gapped-band regions, $E=\sqrt{E_{R}^{2}+E_{I}^{2}}\neq 0,\forall \mathbf{k}\in BZ $, the Jacobian determinant of the transformation
can be expressed as (see Appendix B)
\begin{equation}\label{JD2}
\det J(\mathbf{k},\lambda)=-\frac{\delta (m+\cos k_y)\sin k_x \cos k_y}{E^2}.
\end{equation}

It should be noted that when $\det J(\mathbf{k},\lambda)=0$, the solutions are either $k_{y}= \pm \frac{\pi}{2}$ or $k_{x}=0,\pm \pi$ or $\cos k_{y}=-m$. We find that the quantum states of the zero-Jacobian determinant correspond to the PBS in the complex energy plane. That is why all patterns in the BZ (see Figs.\ref{fig2} and Fig.\ref{fig3}) are horizontal and vertical lines located at either $k_{y}= \pm \frac{\pi}{2}$ or $k_{x}=0,\pm \pi$. The Jacobian determinant is zero for all PBS in the complex energy plane .\cite{Annan2}

The existence of a zero-Jacobian determinant means that the PBS are degenerate in the complex energy plane, which is similar to the previous results obtained for the topological order of the ground states.\cite{Wen} The map between the BZ and complex energy plane is not a homeomorphism.

\section{Conclusions and outlooks}
In summary, as a theoretical model of open and dissipative systems, non-Hermitian quantum systems exhibit rich novel physical phenomena and highlight fundamental issues in condensed matter physics.\cite{Ramy,Gong,Kohei} Non-Hermitian systems contain complex energy structures and their corresponding nonorthogonal eigenstates, which not only lead to novel quantum phases and potential applications, but also a platform for developing new approaches with which explore novel physical properties and mathematical structures beyond those of the canonical Hermitian systems.\cite{Kohei,Annan,Ali}

We have developed a new approach to determine the topological invariants of the quantum states in non-Hermitian systems based on a local-global correspondence between the PBS in the complex energy plane and their corresponding patterns in the BZ. We found that the PBS in the complex energy plane mapped to the BZ forms patterns in the BZ. The patterns are robust due to topological invariants in the gapped regions of the parameter space, which implies a local-global correspondence between the PBS and the topological invariants. We have demonstrated the validity of this approach based on the non-Hermitian Chern insulator model. We predicted the standard phase diagram of the non-Hermitian Chern insulator model, but also uncovered some novel topological invariants embedded in the phase diagram.
The local-global correspondence reveals that only some of quantum states in the complex energy plane dominate the topological invariants of non-Hermitian systems.
In other words, the local information contained in the PBS determines the global property of the quantum states.
This approach provides us with a novel insight into the new physics of non-Hermitian systems beyond that described by the winding number, Chern number and vorticity. This local-global correspondence can be regarded as a generalized bulk-boundary correspondence in the complex energy plane.
Moreover, this local-global correspondence can be expected to work for more generic non-Hermitian models even though we demonstrated it using the specific model, the non-Hermitian Chern insulator model. Specially, it should hold provided that the PBS in the complex energy plane comes from the energy band near the Fermi energy or the Fermi energy surface for strongly interacting non-Hermitian systems.\cite{Wen} Interestingly, the scaling theory of $\mathbb{Z}_2$ topological invariants and universal classes of the topological phase transitions in high-order Dirac models was recently given in.\cite{Chen2} In principle, the local-global correspondence in the complex energy plane can also be generalized to non-Hermitian systems with arbitrary dimensions. For non-Hermitian systems with a small non-Hermitian term, comparative study of differences between these two approaches for arbitrary dimensional systems should be an interesting issue.

To understand the local-global correspondence in the complex energy plane we introduced the generalized vorticity and its corresponding flipping index, with which we demonstrated  the consistency of our results. We also find some connections between the local-global correspondence and the vorticity introduced in previous studies\cite{Ghatak,Fu} The key differences between the local-global correspondence, the generalized vorticity and the Chern number are that the Chern number depends on the exceptional points of the energy band structure whereas the local-global correspondence and generalized vorticity depend on the structure of the PBS in the complex energy plane.

The complex energy plane and its relevant phases play a crucial role for the topological invariants using the local-global correspondence.
The Jacobian determinant of the transformation between the BZ and the complex energy plane is not positive definite, which implies that the BZ and the complex energy plane are not homeomorphic for quantum states.

These results provides some novel insights into the topological invariants of non-Hermitian systems even though there are still questions which require further study,
such as what physical observables emerge from the new topological invariants based on the local-global correspondence and the generalized vorticity, especially what relationships between the local flipping point varying and the global topological invariant of the flipping index and what physical phenomena emergence from the analytic behavior of the imaginary part of the generalized vorticity.
These raise a lot of fundamental issues and their solutions may give rise to practical applications.


\section{Appendixes}

\subsection{Algorithm of local-global correspondence}
\begin{figure}[htbp]
	\centering
	\includegraphics[width=0.7\linewidth]{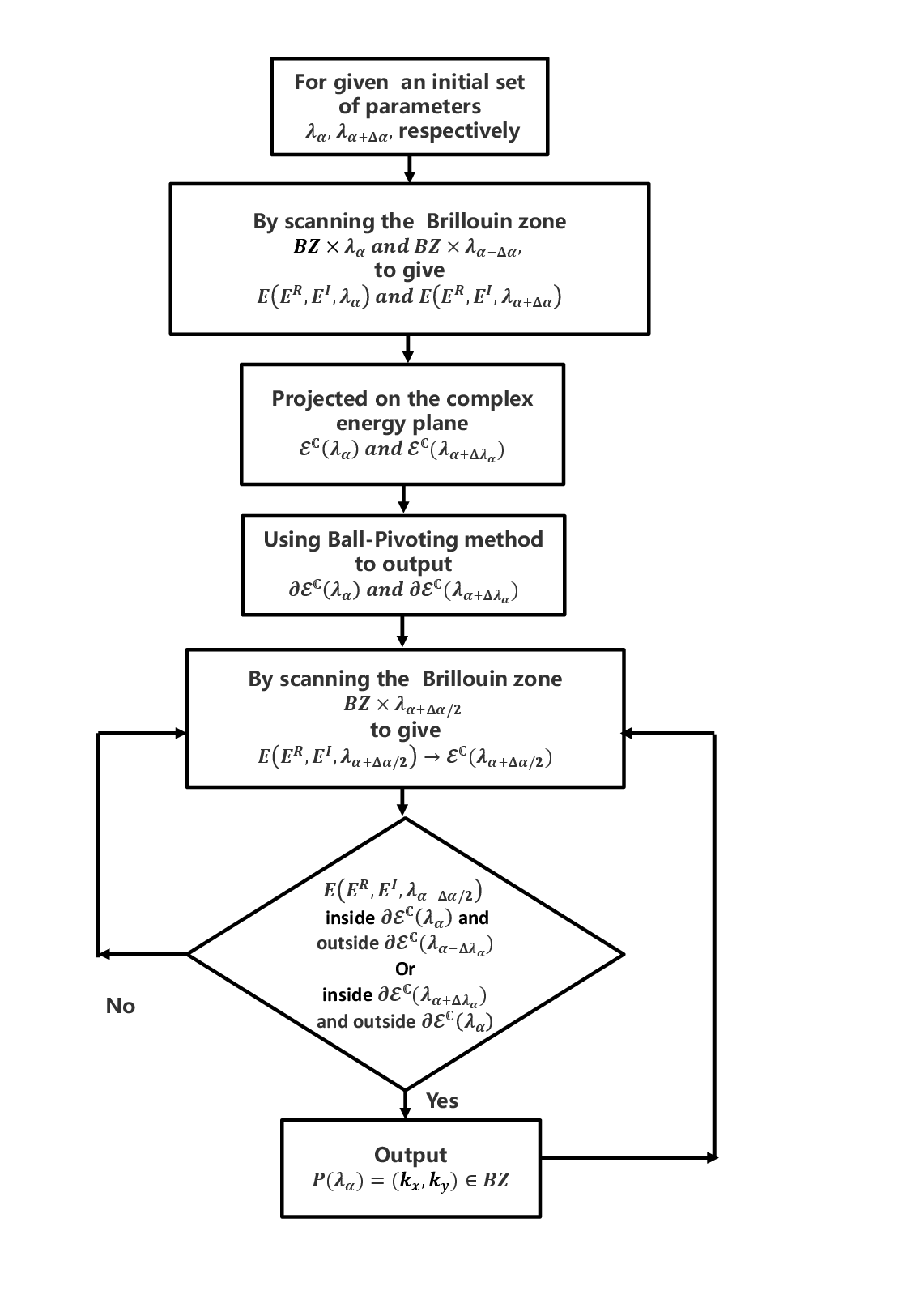}
	\caption{The schematic diagram of the algorithm of the local-global correspondence of the topological phase in the complex energy plane.}
	\label{fig6}
\end{figure}

The local-global correspondence of the topological invariants in the complex energy plane is implemented based on the conceptual and mathematical framework in Sec. II. The numerical method is presented in this section because there is no an analytic method to determine the PBS in the complex energy plane.
The framework of the algorithm to determine the PBS and their corresponding patterns in the BZ is presented in the schematic diagram Fig.\ref{fig6}. The basic steps are as follows.

\begin{itemize}
  \item For given initial set of the parameters $\lambda_\alpha, \lambda_{\alpha+\Delta \alpha}\in \mathcal{M}^2$, where $\lambda_\alpha=(m,\delta)$ in the gapped regions,
  \item we scan the BZ, $(k_{x},k_{y})\in (-\pi,\pi)\times \lambda_\alpha$ and $(k_{x},k_{y})\in (-\pi,\pi)\times \lambda_{\alpha+\Delta \lambda_\alpha}$ and solve the eigen equation of the Hamiltonian to obtain the complex energy bands $E_{\pm}(k_x,k_y; \lambda_\alpha)$ and $E_{\pm}(k_x,k_y; \lambda_{\alpha+\Delta \alpha})$.
      There is an analytic solution for the non-Hermitian Chern insulator model for this step.
  \item The complex energy bands are mapped to the complex energy plane. When we repeat this step for the whole BZ we obtain either one or two regions $\varepsilon^c$ in the complex energy plane,which depend on the parameters. Namely, we obtain two regions of the complex energy band in the complex energy plane for the parameters  $\lambda_\alpha, \lambda_{\alpha+\Delta \alpha}\in \Lambda_{\alpha}\in\mathcal{M}^2$.
  \item Using the ball pivoting algorithm,\cite{Fausto}, we obtain the PBS in the complex energy plane, namely $\partial \mathcal{E}^c(\lambda_\alpha)$ and $\partial\mathcal{E}^c(\lambda_{\alpha+\Delta \lambda_\alpha})$, respectively.
  \item For given an initial set of the parameters between $\lambda_\alpha$ and $\lambda_{\alpha+\Delta \alpha}\in \mathcal{M}^{2}$, namely, $\lambda_{\alpha+\Delta \lambda_\alpha/2}$, we repeat above steps to obtain the PBS, $\partial\mathcal{E}^c(\lambda_{\alpha+\Delta \lambda_\alpha/2})$ in the complex energy plane.
  \item We compare the point $E(E_{R}, E_{I}; \lambda_{\alpha+\Delta\alpha/2})$ in the complex energy plane with the PBS, $\partial\mathcal{E}^c(\lambda_\alpha)$ and $\partial\mathcal{E}^c(\lambda_{\alpha+\Delta \lambda_\alpha})$, when the point $E(E_{R}, E_{I}; \lambda_{\alpha+\Delta\alpha/2})$ is located inside  $\partial\mathcal{E}^c(\lambda_\alpha)$ and outside $\partial\mathcal{E}^c(\lambda_{\alpha+\Delta \lambda_\alpha})$, or conversely, we output the point $E(E_{R}, E_{I}; \lambda_{\alpha+\Delta\alpha/2})$ to the BZ as a PBS in the complex energy plane for the parameter $\lambda_{\alpha}$, which forms a pattern in the BZ. Otherwise, we give up this point, which is not PBS.
  \item By repeating these steps for all BZ (see the schematic diagram in Fig.\ref{fig6}) we obtain the whole pattern of the PBS in the complex energy plane in the BZ.
\end{itemize}

It should be remarked that this numerical algorithm is efficient and stable because the topological phase is robust for the parameter perturbation even though there is some numerical error coming from the ball pivoting algorithm.\cite{Annan2} That is why the patterns contain some flaw in Fig.\ref{fig2} (j)-(m) and Fig.\ref{fig3} (g)-(i). The numerical errors depend on the parameter of the radius of the ball in the ball pivoting algorithm. Reducing the numerical errors costs additional computing time. However, the numerical errors do not change the results because the topological invariant of quantum states is robust for any deformation.

\subsection{Jacobian Determinant}
We present here the derivation of the Jacobian determinant of the transformation between the BZ and the complex energy plane.
Note that the relationship between the real and imaginary parts of the energy band in (\ref{RIEB1}) and (\ref{RIEB2}), is given by
\begin{subequations}\label{ERI2}
\begin{eqnarray}
E^{2}_{R}-E^{2}_{I} &=& \epsilon, \\
E^{2}_{R}+E^{2}_{I} &=& \sqrt{\epsilon^2+\omega^2}, \\
E^{2}_{R}E^{2}_{I} &=&\frac{\omega^{2}}{4}.
\end{eqnarray}
\end{subequations}
Yielding Eqs.(\ref{ERI1}a) and (\ref{ERI1}c), where we set $t=1$ for convenience without losing generality. Taking derivatives with respect to $k_x$ and $k_y$ to (\ref{ERI3}a) and (\ref{ERI3}b), we have
\begin{subequations}\label{DERIx}
\begin{align}
E_{R}\frac{\partial E_R}{\partial k_x}-E_{I}\frac{\partial E_I}{\partial k_x} &= -\sin k_{x}\cos k_{y}-m\sin k_{x}, \\
E_{I}\frac{\partial E_R}{\partial k_x}+E_{R}\frac{\partial E_I}{\partial k_x}&= 0.
\end{align}
\end{subequations}
and
\begin{subequations}\label{DERIy}
\begin{align}
E_{R}\frac{\partial E_R}{\partial k_y}-E_{I}\frac{\partial E_I}{\partial k_y} &= -\cos k_{x}\sin k_{y}-m\sin k_{y}, \\
E_{I}\frac{\partial E_R}{\partial k_y}+E_{R}\frac{\partial E_I}{\partial k_y} &= \delta \cos k_{y}.
\end{align}
\end{subequations}
We can rewrite (\ref{DERIx}) and (\ref{DERIy}) to give
\begin{equation}\label{DERIx2}
\left(
\begin{array}{cc}
E_{R} & -E_{I} \\
E_{I} & E_{R}
\end{array}
\right)
\left(
\begin{array}{c}
\frac{\partial E_R}{\partial k_x} \\
\frac{\partial E_I}{\partial k_x}
\end{array}
\right)=\left(
\begin{array}{c}
-(\cos k_{y}+m)\sin k_{x} \\
0
\end{array}
\right),
\end{equation}
and
\begin{equation}\label{DERIy2}
\left(
\begin{array}{cc}
E_{R} & -E_{I} \\
E_{I} & E_{R}
\end{array}
\right)
\left(
\begin{array}{c}
\frac{\partial E_R}{\partial k_y} \\
\frac{\partial E_I}{\partial k_y}
\end{array}
\right)=\left(
\begin{array}{c}
-(\cos k_{x}+m)\sin k_{y} \\
\delta\cos k_y
\end{array}
\right).
\end{equation}
For the gapped-band regions, $E=\sqrt{E_{R}^{2}+E_{I}^{2}}\neq 0,\forall \mathbf{k}\in BZ $, there exists the inverse of the energy matrix in (\ref{DERIx2}) and(\ref{DERIy2}). Thus, the derivatives of the real and imaginary energy bands can be expressed as
\begin{equation}\label{DERIx3}
\left(
\begin{array}{c}
\frac{\partial E_R}{\partial k_x} \\
\frac{\partial E_I}{\partial k_x}
\end{array}
\right)=
\left(
\begin{array}{cc}
E_{R} & -E_{I} \\
E_{I} & E_{R}
\end{array}
\right)^{-1}
\left(
\begin{array}{c}
-(\cos k_{y}+m)\sin k_{x} \\
0
\end{array}
\right),
\end{equation}
and
\begin{equation}\label{DERIy4}
\left(
\begin{array}{c}
\frac{\partial E_R}{\partial k_y} \\
\frac{\partial E_I}{\partial k_y}
\end{array}
\right)=
\left(
\begin{array}{cc}
E_{R} & -E_{I} \\
E_{I} & E_{R}
\end{array}
\right)^{-1}
\left(
\begin{array}{c}
-(\cos k_{x}+m)\sin k_{y} \\
\delta\cos k_y
\end{array}
\right),
\end{equation}
where the inverse of the matrix is given by
\begin{equation}\label{DERIy2}
\left(
\begin{array}{cc}
E_{R} & -E_{I} \\
E_{I} & E_{R}
\end{array}
\right)^{-1}=\frac{1}{E^{2}}
\left(
\begin{array}{cc}
E_{R} & E_{I} \\
-E_{I} & E_{R}
\end{array}
\right),
\end{equation}
with $E=\sqrt{E^{2}_{R}+E^{2}_{I}}$.
Consequently, the derivatives of the real and imaginary parts of the energy bands with respect to $k_x$ and $k_y$ are given by
\begin{equation}\label{DERIx5}
\left(
\begin{array}{c}
\frac{\partial E_R}{\partial k_x} \\
\frac{\partial E_I}{\partial k_x}
\end{array}
\right)=\frac{1}{E^{2}}
\left(
\begin{array}{c}
-E_{R}(\cos k_{y}+m)\sin k_{x} \\
E_{I}(\cos k_{y}+m)\sin k_{x}
\end{array}
\right),
\end{equation}
and
\begin{equation}\label{DERIy6}
\left(
\begin{array}{c}
\frac{\partial E_R}{\partial k_y} \\
\frac{\partial E_I}{\partial k_y}
\end{array}
\right)=\frac{1}{E^{2}}
\left(
\begin{array}{c}
\delta E_{I}\cos k_{y}-E_{R}(\cos k_x+m)\sin k_{y} \\
\delta E_{R}\cos k_{y}+E_{I}(\cos k_x+m)\sin k_{y}
\end{array}
\right).
\end{equation}

The Jacobian determinant of the transformation given in (\ref{JD1})
can be expressed as
\begin{equation}\label{JD2}
\det J(\mathbf{k},\lambda)=
\begin{vmatrix}
\frac{\partial E_{R}}{\partial k_{x}} & \frac{\partial E_{R}}{\partial k_{y}} \\
\frac{\partial E_{I}}{\partial k_{x}} & \frac{\partial E_{I}}{\partial k_{y}}
\end{vmatrix}.
\end{equation}
By substituting (\ref{DERIx5}) and (\ref{DERIy6}) into (\ref{JD2}), we obtain the Jacobian determinant of the transformation from the BZ to the complex energy plane
\begin{equation}\label{JD3}
\det J(\mathbf{k},\lambda)=-\frac{\delta (m+\cos k_y)\sin k_x \cos k_y}{E^2}.
\end{equation}

\bibliography{apssamp}

\end{document}